\documentclass[review,onefignum,onetabnum]{article}
\pdfoutput=1

\usepackage{latexsym}
\usepackage{amssymb,amsbsy,amsmath,amsfonts,amssymb,amscd,mathtools,amsfonts,mathrsfs,graphicx,color,dsfont,pdfsync,bbm}

\newcommand {\Chi} {{\bf \raise 2pt \hbox{$\chi$}} }

\newcommand {\sgn} { {\rm sgn} }

\newcommand {\rmd} { {\mathrm d} }
\newcommand{\beq}{\begin{equation}}
\newcommand{\eeq}{\end{equation}}
\newcommand{\beqa}{\begin{eqnarray}}
\newcommand{\eeqa}{\end{eqnarray}}
\newcommand{\bea} {\begin{array}{rl}}
\newcommand{\eea} {\end{array}}
\newcommand{\bepa}{\left\{ \begin{array}{l}}
\newcommand{\eepa} {\end{array}\right.}

\title{Discrete and continuum models for the evolutionary and spatial dynamics of cancer: a very short introduction through two case studies}

\author{Tommaso Lorenzi\thanks{University of St Andrews, UK 
  (tl47@st-andrews.ac.uk)}
\and
Fiona R. Macfarlane\thanks{University of St Andrews, UK 
  (frm3@st-andrews.ac.uk)}
\and
Chiara Villa\thanks{University of St Andrews, UK 
  (cv23@st-andrews.ac.uk)}
}

\begin{document}
\date{}
\maketitle

\begin{abstract}
We give a very short introduction to discrete and continuum models for the evolutionary and spatial dynamics of cancer through two case studies: a model for the evolutionary dynamics of cancer cells under cytotoxic therapy and a model for the mechanical interaction between healthy and cancer cells during tumour growth. First we develop the discrete models, whereby the dynamics of single cells are described through a set of rules that result in branching random walks. Then we present the corresponding continuum models, which are formulated in terms of non-local and nonlinear partial differential equations, and we summarise the key properties of their solutions. Finally, we carry out numerical simulations of the discrete models and we construct numerical solutions of the corresponding continuum models. The biological implications of the results obtained are briefly discussed.
\end{abstract}

\section{Introduction}
Mathematical models have been increasingly used to dissect the variety of complex processes that orchestrate the evolutionary and spatial dynamics of cancer~\cite{altrock2015mathematics,anderson2008integrative,Anderson2018,Beerenwinkel2016,byrne2010dissecting,chisholm2016cell,gatenby2003mathematical}. In particular, integro-differential equations and nonlocal partial differential equations (PDEs) for the dynamics of cell population densities (\emph{i.e.} phenotypic distributions of cell populations) have helped to elucidate the adaptive processes that are at the root of cancer resistance to cytotoxic agents~\cite{almeida2018evolution,chisholm2015emergence,cho2017modeling,cho2018modeling,delitala2012mathematical,lavi2013role,lavi2014simplifying,lorenzi2016tracking,lorenzi2018role,lorz2015modeling,Lorz:2013,pouchol2018asymptotic}. Furthermore, nonlinear PDEs for the evolution of cellular densities in response to pressure gradients generated by cell proliferation have shed light on the underpinnings of tumour growth and cancer invasion~\cite{ambrosi2002mechanics,ambrosi2002closure,araujo2004history,bresch2010computational,byrne2010dissecting,byrne1995growth,byrne1996growth,byrne1997free,byrne2009individual,byrne2003two,byrne2003modelling,chaplain2006mathematical,chen2001influence,ciarletta2011radial,greenspan1976growth,lowengrub2009nonlinear,perthame2014some,preziosi2003cancer,ranft2010fluidization,roose2007mathematical,sherratt2001new,ward1999mathematical,ward1997mathematical}.
 
PDE models for the evolutionary and spatial dynamics of cancer are amenable to both numerical and analytical approaches, which support an in-depth theoretical understanding of the application problems under study. However, defining these models on the basis of phenomenological considerations can hinder a precise mathematical representation of key biological aspects. Therefore, it is desirable to derive them from first principles as the appropriate continuum limit of discrete models that track the dynamics of single cells ({\it i.e.} individual-based models). In fact, since individual-based models enable a more direct representation of fine details of cell dynamics, this ensures that key biological aspects are faithfully mirrored in the structure of the PDEs considered. As a result, the derivation of continuum models for the evolutionary~\cite{champagnat2002canonical,champagnat2006unifying,champagnat2007invasion,chisholm2016evolutionary,stace2019} and spatial~\cite{baker2019free,binder2009exclusion,buttenschoen2018space,byrne2009individual,chaplain2019bridging,deroulers2009modeling,drasdo2005coarse,dyson2012macroscopic,fernando2010nonlinear,othmer2000diffusion,hillen2009user,inoue1991derivation,johnston2017co,johnston2012mean,landman2011myopic,lushnikov2008macroscopic,motsch2018short,murray2009discrete,murray2012classifying,oelschlager1989derivation,oelschlager1990large,othmer1988models,Painter2002,painter2003modelling,penington2011building,penington2014interacting,simpson2010cell,simpson2007simulating,stevens2000derivation,stevens1997aggregation} dynamics of cell populations from underlying individual-based models has become an active research area.

In this paper, we give a very short introduction to such discrete and continuum models for the evolutionary and spatial dynamics of cancer through two case studies. In Section~\ref{Section1}, we develop a stochastic individual-based model for the evolutionary dynamics of cancer cells under cytotoxic therapy, and we introduce the corresponding deterministic continuum model, which is formulated in terms of a non-local PDE for the cell population density. In Section~\ref{Section2}, we present a stochastic individual-based model for the mechanical interaction between healthy and cancer cells during tumour growth, and we discuss the corresponding deterministic continuum model, which comprises a system of coupled nonlinear PDEs for the cell densities. We summarise some properties of the solutions of the PDEs, we carry out a quantitative comparison between their numerical solutions and the results of numerical simulations of the corresponding individual-based models, and we infer the key biological implications of the results obtained. In Section~\ref{Section4}, we provide a brief overview of possible developments of the models that could capture additional layers of biological complexity.

\section{Discrete and continuum models for the evolutionary dynamics of cancer cells under cytotoxic therapy}
\label{Section1}
In this section, we use the modelling framework developed by Stace {\it et al.}~\cite{stace2019} to define a stochastic individual-based model for the phenotypic evolution of cancer cells under cytotoxic therapy (Section~\ref{sec:discreteEvoDyn}). Moreover, we present a non-local PDE for the cell population density that can be formally derived from this discrete model by passing to the continuum limit, and we summarise some key properties of its solutions (Section~\ref{sec:cont1}). Finally, we carry out a quantitative comparison between the results of numerical simulations of the individual-based model and numerical solutions of the corresponding PDE (Section~\ref{secnum1n}).

\subsection{Discrete model}
\label{sec:discreteEvoDyn}
We consider a population of cancer cells exposed to the action of a cytotoxic agent. Cells within the population proliferate (\emph{i.e.} divide and die) and undergo heritable, spontaneous phenotypic variations. The phenotypic state of every cell is characterised by a variable $y \in \mathbb{R}$, which represents the rescaled level of expression of a gene that controls both cell proliferation and cytotoxic-drug resistance~\cite{hanahan2011hallmarks,medema2013cancer}. On the basis of previous experimental and theoretical studies~\cite{gatenby2009adaptive,pisco2015non,silva2012evolutionary}, we assume that there is a sufficiently high level of gene expression $y^*$ which makes the cells fully resistant to the cytotoxic agent and a sufficiently low level of gene expression $y_*<y^*$ conferring the highest rate of cellular division. Without loss of generality, we define $y^* := 1$ and $y_* := 0$. 

We represent each cell as an agent that occupies a position on a lattice, and we model cell proliferation and heritable, spontaneous phenotypic variations according to a set of simple rules that result in a discrete-time branching random walk. We discretise the time variable $t \in \mathbb{R}_{\geq 0}$ and the phenotypic state $y$ as $t_{k} = k \tau$ with $k\in\mathbb{N}_{0}$ and $y_{i} = i  \chi$ with $i \in \mathbb{Z}$, respectively, where $0<\tau, \chi \ll 1$. We introduce the dependent variable $N^{k}_{i}\in\mathbb{N}_0$ to model the number of cells on the lattice site $i$ at the time-step $k$, and we define the cell population density and the size of the cell population (\emph{i.e.} the total number of cells), respectively, as
\begin{equation}
\label{e:SP1}
n(t_k,y_{i})= n^k_{i} := N^{k}_{i} \,  \chi^{-1}, \quad  \rho(t_k)=\rho^k := \sum_i N^{k}_{i}.
\end{equation}
Moreover, we define the mean phenotypic state and the related standard deviation, respectively, as
\begin{equation}
\label{e:SP2}
\mu(t_k) = \mu^k := \frac{1}{\rho^k} \sum_{i} y_i \, N^{k}_i, \quad \sigma(t_k) = \sigma^{k} := \sqrt{\frac{1}{\rho^k} \sum_{i} y^2_i \, N^{k}_i  -  \left(\mu^k\right)^2}.
\end{equation}
The standard deviation $\sigma^{k}$ provides a possible measure of the level of phenotypic heterogeneity within the cell population at the $k^{th}$ time-step.
\newpage
The dynamic of cancer cells is governed by the following rules.
\\\\
{\bf Mathematical modelling of heritable, spontaneous phenotypic variations} We model the effect of heritable, spontaneous phenotypic variations by allowing cancer cells to update their phenotypic states according to a random walk. In particular, at each time-step $k$ every cell in the population can enter into a new phenotypic state with probability $\lambda \in [0,1]$, or remain in its current phenotypic state with probability $1-\lambda$. A focal cell in the phenotypic state $y_i$ that undergoes a phenotypic variation can enter either into the phenotypic state $y_{i-1}$ or into the phenotypic state $y_{i+1}$ with probability $\lambda/2$.
\\\\
{\bf Mathematical modelling of cell proliferation} We allow cancer cells to divide, die or remain quiescent with probabilities that depend on their phenotypic states, as well as on the environmental conditions given by the size of the cell population and the concentration of the cytotoxic agent. We assume that a dividing cell is replaced by two identical progeny cells that inherit the phenotypic state of the parent cell (\emph{i.e.} the progeny cells are placed on the original lattice site of the parent cell). 

We denote by $b(y_i)$ the net division rate of a focal cell in the phenotypic state $y_i$ (\emph{i.e.} the difference between the rate of cell division and the rate of apoptosis). To take into account the fact that the phenotypic state $y=0$ corresponds to the highest rate of cell division, we let the net cell division rate $b: \mathbb{R} \to \mathbb{R}$ satisfy following assumptions
\begin{equation}\label{a.p}
b(0) > 0, \quad b'(0) = 0 \quad \text{and} \quad b''(\cdot) < 0.
\end{equation}
The fact that the net proliferation rate $b(x)$ can become negative for values of $y$ sufficiently far from the maximum point $y=0$ models the fact that unfit phenotypic variants cannot survive within the population.

Moreover, to translate into mathematical terms the idea that higher cell numbers correspond to less available space and resources, and thus to more intense intrapopulation competition, at every time-step $k$ we allow cancer cells to die due to intrapopulation competition at rate $d(\rho^k)$, where the function $d : \mathbb{R}_{\geq 0} \to \mathbb{R}_{\geq 0}$ satisfies the following assumptions
\begin{equation}\label{a.d}
d(0)=0 \quad \text{and} \quad d'(\cdot)>0.
\end{equation}

Finally, we denote by $\kappa(y_i,c^k)$ the rate at which a focal cell in the phenotypic state $y_i$ can be induced to death by the rescaled concentration $c(t_k) = c^k$ of the cytotoxic agent, with $c : \mathbb{R}_{\geq 0} \to \mathbb{R}_{\geq 0}$. Since cells in the phenotypic state $y=1$ are fully resistant to the cytotoxic agent and, for cells in phenotypic states other than $y=1$, the rate of death induced by the cytotoxic agent increases with the dose of the agent, we assume that the function $\kappa: \mathbb{R} \times \mathbb{R}_{\geq 0} \to \mathbb{R}_{\geq 0}$ satisfies the following conditions
\beq
\label{a.k1}
\kappa(1,c) = 0, \quad  \partial_{y} \kappa(1,c) = 0 \quad \text{and} \quad \partial^2_{yy} \kappa(\cdot,c) > 0 \;\; \forall \, c > 0
\eeq
and
\beq
\label{a.k2}
\kappa(\cdot,0) = 0, \quad \partial_c \kappa(y,\cdot) \geq 0 \;\; \forall \, y \neq 1.
\eeq

Therefore, at the $k^{th}$ time-step a focal cell on the lattice site $i$ can divide with probability 
\begin{equation}\label{a.b1}
\tau \, b(y_i)_+ \quad \text{where} \quad b(y_i)_+ = \max \left(0, b(y_i)\right),
\end{equation}
or die with probability 
\begin{equation}\label{a.b2}
\tau \, \big(b(y_i)_- + d(\rho^k) + \kappa(y_i,c^k \big) \quad \text{where} \quad b(y_i)_-=- \min \left(0, b(y_i)\right)
\end{equation}
or remain quiescent with probability 
\begin{equation}\label{a.b3}
1 - \tau \, \big(\left|b(y_i)\right| + d(\rho^k) + \kappa(y_i,c^k) \big) \quad \text{where} \quad \left|b(y_i)\right| = b(y_i)_+ + b(y_i)_-.
\end{equation}
We assume the time-step $\tau$ to be sufficiently small so that the quantities~\eqref{a.b1}-\eqref{a.b3} are all between 0 and 1. In this mathematical framework, the fitness of a focal cell in the phenotypic state $i$ at the time-step $k$ under the environmental conditions determined by the population size $\rho^k$ and the concentration of the cytotoxic agent $c^k$ (\emph{i.e.} the phenotypic fitness landscape of the cancer cell population~\cite{huang2013genetic,merlo2006cancer,poelwijk2007empirical}) is defined as
\beq
\label{def:fitness}
R(y_i,\rho^k,c^k) := b(y_i) - d(\rho^k) - \kappa(y_i,c^k).
\eeq

Following Lorenzi \emph{et al.}~\cite{lorenzi2016tracking}, among the possible definitions of the functions $b(y)$, $d(\rho)$ and $\kappa(y,c)$ that satisfy assumptions~\eqref{a.p}-\eqref{a.k2} we consider
\begin{equation}
\label{e.bdk}
b(y) := \gamma - \eta \, y^2, \quad  d(\rho) := \zeta \, \rho, \quad \kappa(y,c) :=  c \, (1-y)^2.
\end{equation}
In~\eqref{e.bdk}, the parameter $\gamma \in \mathbb{R}_{> 0}$ is the division rate of the fastest dividing cells in the phenotypic state $y=0$, while the parameter $\eta \in \mathbb{R}_{> 0}$ is a nonlinear selection gradient that provides a measure of the strength of natural selection when the cytotoxic agent is not present. Finally, the parameter $\zeta \in \mathbb{R}_{> 0}$ is inversely proportional to the carrying capacity of the cancer cell population. Definitions~\eqref{e.bdk} satisfy assumptions~\eqref{a.p}-\eqref{a.k2} and ensure analytical tractability of the deterministic continuum counterpart of the model, which is presented in the next section. Moreover, these definitions lead to a fitness function $R(y,\rho,c)$ that is close to the approximate fitness landscapes which can be inferred from experimental data through regression techniques~\cite{otwinowski2014inferring}. In fact, substituting~\eqref{e.bdk} into~\eqref{def:fitness}, a little algebra shows that
\beq
\label{def:fitnessspec}
R\big(y,\rho,c\big) = r_0(c) - r_1(c) \left(y - Y(c) \right)^2 - \zeta \, \rho
\eeq
with
\beq
\label{def:ror1yfit}
r_0(c) := \gamma - \frac{\eta \, c}{\eta+c}, \quad r_1(c) := \eta + c, \quad  Y(c) : = \frac{c}{\eta + c},
\eeq
where $r_0(c)$ is the maximum fitness, $Y(c)$ is the fittest phenotypic state and $r_1(c)$ is a nonlinear selection gradient.

\subsection{Corresponding continuum model and properties of its solutions}
\label{sec:cont1}
Using the formal method presented in~\cite{chisholm2016evolutionary,stace2019} and letting $\tau \to 0$ and $\chi \to 0$ in such a way that
\beq
\label{asymatdr}
\frac{\lambda \chi^2}{2 \tau} \rightarrow \beta \in \mathbb{R}_{>0},
\eeq
where the parameter $\beta$ is the rate of heritable, spontaneous phenotypic variations, it is possible to show that the deterministic continuum counterpart of the stochastic individual-based model presented in Section~\ref{sec:discreteEvoDyn} is given by the following conservation equation for the population density function of cancer cells $n(t,y) \geq 0$:
\begin{equation}
\label{e.PDE}
\begin{cases}
\displaystyle{\partial_t n  - \beta \; \partial^2_{yy} n  =  R(y,\rho(t),c(t)) \, n}, \\\\
\displaystyle{\rho(t) := \int_{\mathbb{R}} n(t,y) \, {\rm d}y},
\end{cases}
\quad (t,y) \in (0, \infty) \times \mathbb{R}.
\end{equation}
In the non-local PDE \eqref{e.PDE}, the function $R(y,\rho,c)$ is defined according to~\eqref{def:fitness}.  

Considering the case where the fitness function $R(y,\rho,c)$ is of the form~\eqref{def:fitnessspec} and the concentration of cytotoxic agent is constant (\emph{i.e.} $c(t) \equiv C \geq 0$), using the method of proof developed in~\cite{chisholm2016evolutionary,lorenzi2015dissecting}, Lorenzi \emph{et al.}~\cite{lorenzi2016tracking} proved that the solution to the Cauchy problem defined by~\eqref{e.PDE} subject to some non-negative and sufficiently regular initial condition is such that
\begin{equation}\label{a0}
\rho(t) \xrightarrow[t  \rightarrow \infty]{}  \max \left(0,\overline{\rho}_C\right) \quad \text{with} \quad \overline{\rho}_C = \frac{1}{\zeta}\left(r_0(C) - \sqrt{\beta \, r_1(C)}\right).
\end{equation}
Moreover, if $\overline{\rho}_C > 0$ there exists a unique non-negative, non-trivial steady-state solution $\overline{n}_C(x)$ of~\eqref{e.PDE}, which is of the Gaussian form
\begin{equation}\label{e.barn}
\overline{n}_C(y) = \frac{\overline{\rho}_C}{\sqrt{2 \pi \overline{\sigma}_C^2}} \, \exp\left[- \frac{1}{2} \, \frac{\left(y - \overline{\mu}_C \right)^2}{\overline{\sigma}_C^2} \right]
\end{equation}
with
\begin{equation}\label{e.barmusigma}
\overline{\mu}_C = Y(C) \quad \text{and} \quad \overline{\sigma}_C^2 = \sqrt{\frac{\beta}{r_1(C)}}.
\end{equation}

\subsection{Quantitative comparison between discrete and continuum models}
\label{secnum1n}  
In this section, we compare the outcomes of numerical simulations of the stochastic individual-based model presented in Section~\ref{sec:discreteEvoDyn} with numerical solutions of the deterministic continuum model given by~\eqref{e.PDE}, and we briefly discuss their biological implications. The results obtained indicate excellent agreement between the simulation results for the individual-based model, the numerical solutions of the continuum model and the long-time asymptotic results summarised in Section~\ref{sec:cont1}.

\subsubsection{Numerical methods and set-up of numerical simulations}
To construct numerical solutions of~\eqref{e.PDE}, we use a uniform discretisation consisting of 1200 points on the interval $[-4,4]$ as the computational domain of the independent variable $y$. We employ a three-point finite difference explicit scheme for the diffusion term and an explicit finite difference scheme for the reaction term~\cite{leveque2007finite} to solve numerically~\eqref{e.PDE} subject to no-flux boundary conditions and to the following initial condition
\beq
\label{ic111pdeevodyn}
n(0,y) := a_0 \exp\left[- a_1 (y - a_2)^2 \right]
\eeq
where
$$
a_0 = 2.63 \times 10^4, \quad a_1 = 39 , \quad a_2 = 0.5.
$$
In agreement with previous papers~\cite{bozic2012dynamics,bozic2013evolutionary,lorenzi2016tracking,Pisco:2013,steel1966growth}, we define 
\begin{equation}\label{paramval1}
\gamma := 0.6, \quad \eta := 0.3, \quad \zeta := 6 \times 10^{-5}, \quad \beta:=5 \times 10^{-3}
\end{equation}
and we consider the cytotoxic agent concentration to be constant, that is, we assume $c \equiv C$ with $C$ expressed in terms of the $LD\alpha$, \emph{i.e.}  the constant value of $c$ that is required to reduce the equilibrium value of the population size $\rho$ by the $\alpha\%$. The parameter values~\eqref{paramval1} along with the values of $C$ considered here are such that the cell population size $\overline{\rho}_C$ given by~\eqref{a0} is positive.

An analogous set-up is used to carry out computational simulations of the discrete model with $\tau := 10^{-3}$, $\chi := 10^{-2}$ and $\lambda$ such that condition~\eqref{asymatdr} is met. At each time-step, we follow the procedures summarised hereafter to simulate heritable, spontaneous phenotypic variations and cell proliferation. Numerical simulations are performed in {\sc Matlab} and all random numbers mentioned below are real numbers drawn from the standard uniform distribution on the interval $(0,1)$ using the built-in function {\sc rand}. 
\\\\
{\bf Numerical simulation of heritable, spontaneous phenotypic variations} For each cell, a random number is generated and it is determined whether or not the cell undergoes a phenotypic variation by comparing this number with the value of the probability $\lambda$. If a cell undergoes a phenotypic variation, a new random number is generated and if the number is less than or equal to $\lambda/2$ then the cell will move into the phenotypic state to the left of its current state, whereas if the number is greater than $\lambda/2$ then the cell will move into the phenotypic state to the right of its current state. No-flux boundary conditions are implemented by letting the attempted phenotypic variation of a cell be aborted if it requires moving into a phenotypic state that does not belong to the computational domain.
\\\\
{\bf Numerical simulation of cell proliferation} The size of the cell population is computed and the probabilities of cell division, death and quiescence are evaluated for every phenotypic state via~\eqref{a.b1}-\eqref{a.b3} and \eqref{e.bdk}. For each cell, a random number is generated and the cells' fate is determined by comparing this number with the probabilities of division, death and quiescence corresponding to the cell phenotypic state.
                                                                           
\subsubsection{Main results}
\begin{figure}[h!]
\centering
\includegraphics[width=0.78\textwidth]{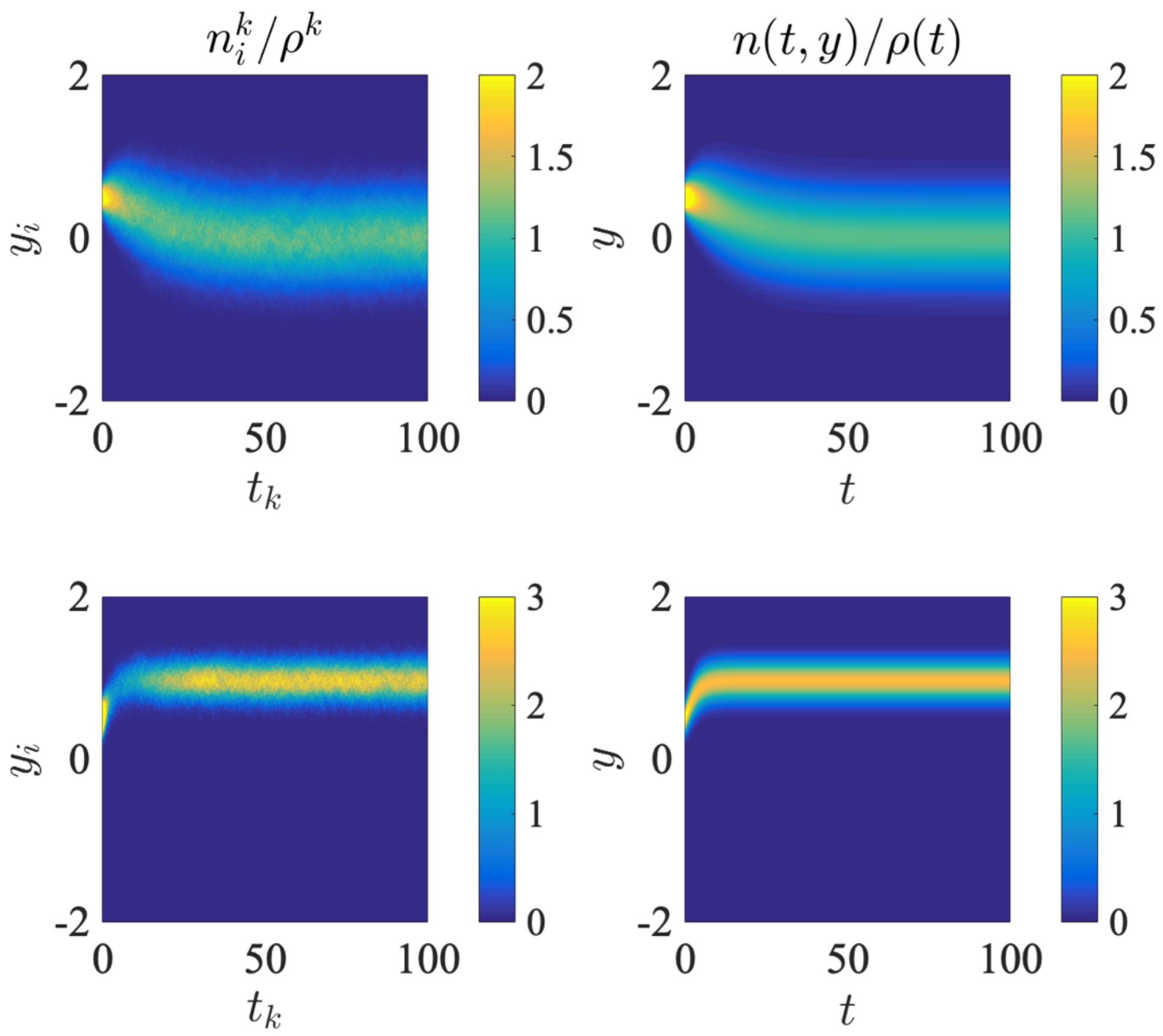}
\caption{{\bf Dynamics of the cell population density.} Dynamics of the solution of~\eqref{e.PDE} (right panels) and the cell population density computed from numerical simulations of the individual-based model (left panels) for $C=0$ (top panels) and $C>0$ corresponding to the $LD80$ (bottom panels). 
}
\label{fig0}
\end{figure}

\begin{figure}[h!]
\centering
\includegraphics[width=0.85\textwidth]{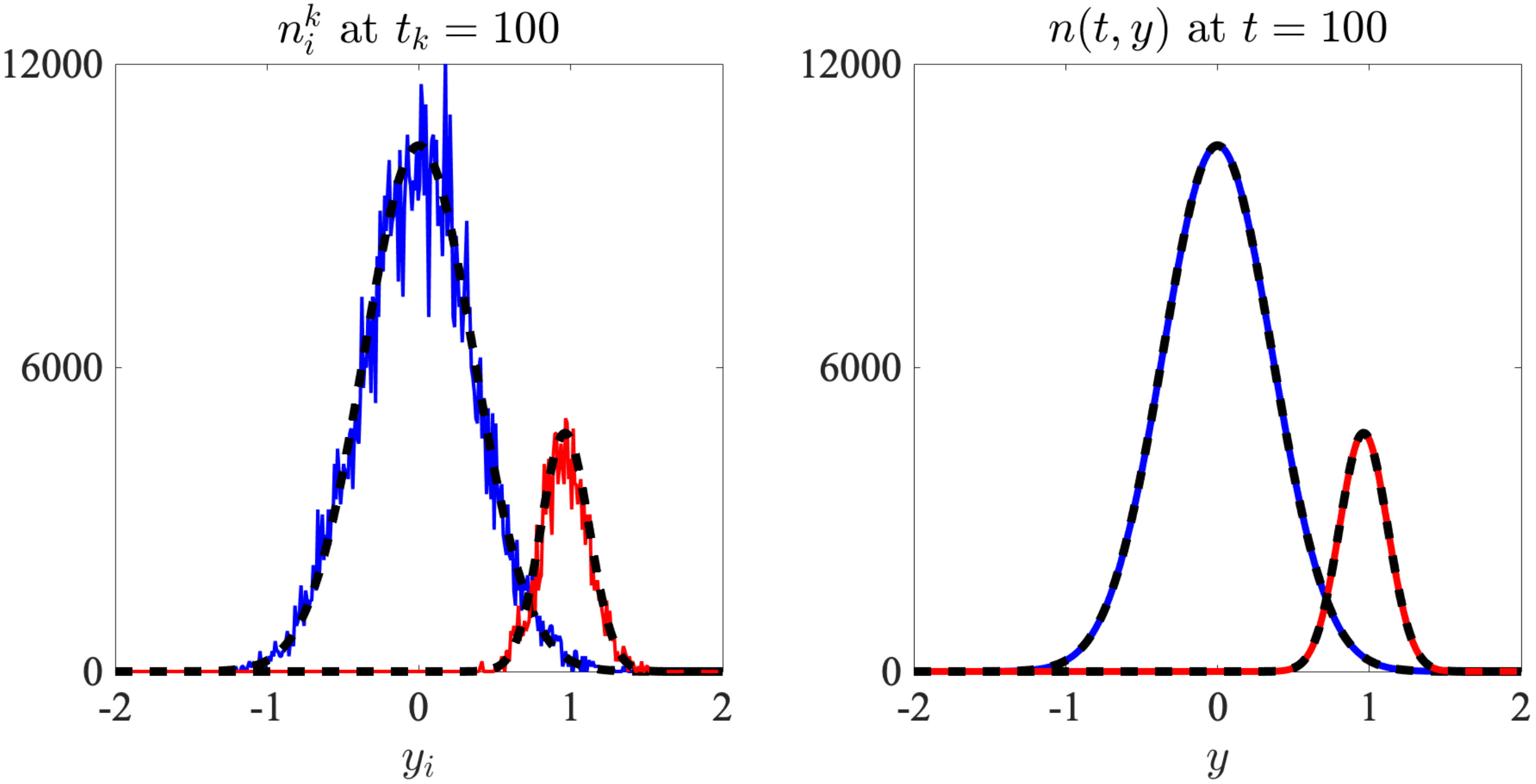}
\caption{{\bf Cell population density at the end of numerical simulations.} Plots of the solution of~\eqref{e.PDE} at the final time of simulations $t=100$ (right panel) and of the cell population density computed at the end of numerical simulations of the individual-based model (left panel) for $C=0$ (blue lines) and $C>0$ corresponding to the $LD80$ (red lines). The black dashed lines highlight the steady-state solution~\eqref{e.barn}. }
\label{fig1}
\end{figure}

\begin{figure}[h!]
\centering
\includegraphics[width=\textwidth]{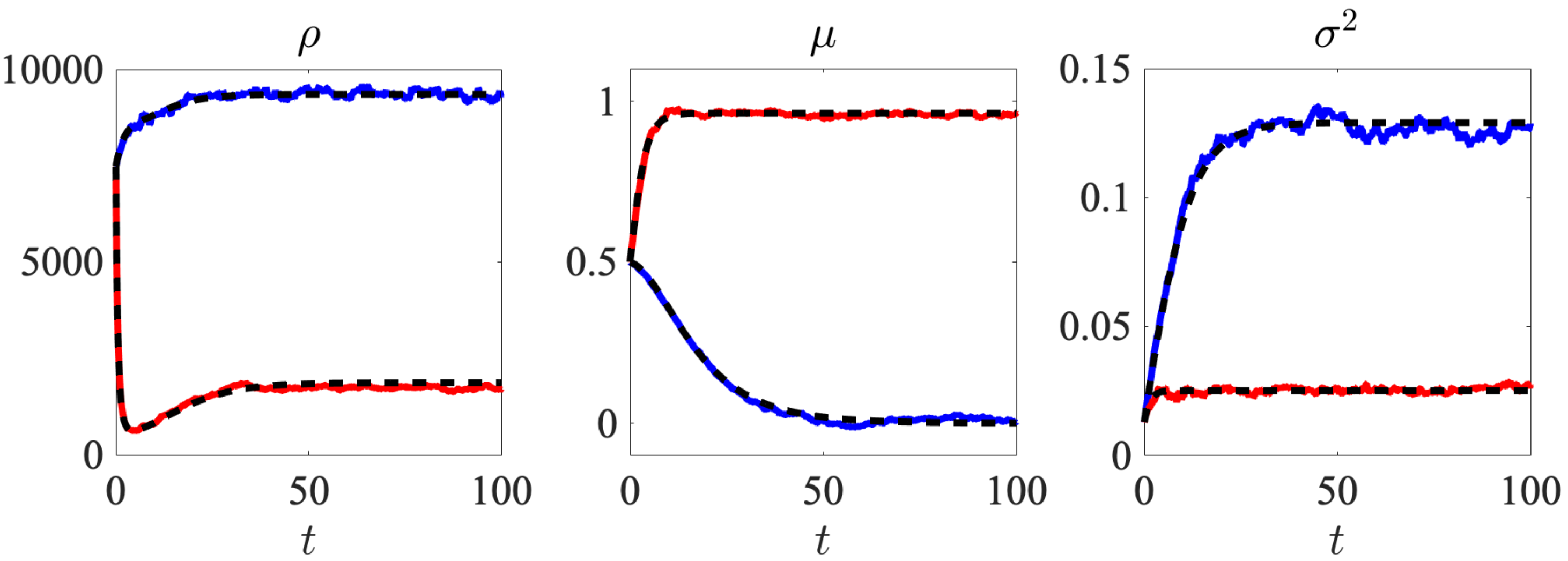}
\caption{{\bf Dynamics of the population size, mean phenotypic state and related variance.} Dynamics of the population size (left panel), mean phenotypic state (central panel) and related variance (right panel) corresponding to the cell population densities of Figure~\ref{fig0}. Solid lines highlight quantities computed from the results of numerical simulations of the individual-based model, while dashed lines highlight quantities computed from numerical solutions of~\eqref{e.PDE}. Blue and red lines refer to the case where $C=0$ and $C>0$ corresponding to the $LD80$, respectively.
}
\label{fig2}
\end{figure}

The plots in Figure~\ref{fig0} show sample dynamics of the cell population density computed from numerical simulations of the individual-based model (left panels) and the solution of~\eqref{e.PDE} (right panels) for $C=0$ (top panels) and $C>0$ corresponding to the $LD80$ (bottom panels). Moreover, Figure~\ref{fig1} shows a comparison between the cell population density computed at the end of numerical simulations of the individual-based model, or the solution of~\eqref{e.PDE} at the final time of simulations $t=100$, and the steady-state solution~\eqref{e.barn}. Finally, Figure~\ref{fig2} displays the dynamics of the corresponding population size (left panel), mean phenotypic state (central panel) and related variance (right panel) computed from numerical simulations of the individual-based model (solid lines) and from numerical solutions of~\eqref{e.PDE} (dashed lines). 

In agreement with the asymptotic results summarised in Section~\ref{sec:cont1}, the numerical results presented in Figures~\ref{fig0} and~\ref{fig1} demonstrate that the cell population density converges to the steady-state solution $\overline{n}_C(y)$ given by~\eqref{e.barn}. Accordingly, the numerical results presented in Figure~\ref{fig2} show that the population size, the mean phenotypic state and the related variance converge, respectively, to the asymptotic values $\overline{\rho}_C$, $\overline{\mu}_C$ and $\overline{\sigma}_C^2$ given by~\eqref{a0} and~\eqref{e.barmusigma}.   
\\\\
\noindent{\bf Biological implications in brief} The results summarised by Figures~\ref{fig0}-\ref{fig2} communicate the biological notion that when the cytotoxic agent is not administered the population evolves to be mainly composed of highly proliferative phenotypic variants (\emph{i.e.} cells in phenotypic states close to $y=0$). On the other hand, administering the cytotoxic agent leads to a population bottleneck, resulting in a reduced size of the cancer cell population and a lower level of phenotypic heterogeneity, at the cost of promoting the selection of highly resistant phenotypic variants (\emph{i.e.} cells in phenotypic states close to $y=1$). 

\section{Discrete and continuum models for the mechanical interaction between healthy and cancer cells during tumour growth}
\label{Section2}
In this section, we use the modelling framework developed by Chaplain {\it et al.}~\cite{chaplain2019bridging} to define a stochastic individual-based model for the mechanical interaction between healthy and cancer cells during tumour growth (Section~\ref{sec:discreteTumInv}). Moreover, we present a system of coupled nonlinear PDEs for the densities of healthy and cancer cells that can be formally derived from this discrete model by passing to the continuum limit, and we summarise some key properties of its solutions (Section~\ref{contmod2}). Finally, we carry out a quantitative comparison between the results of numerical simulations of the individual-based model and numerical solutions of the corresponding PDE (Section~\ref{secnum2n}).
\newpage

\subsection{Discrete model}
\label{sec:discreteTumInv}
We consider a one dimensional section of a normal tissue composed of healthy cells in homeostatic equilibrium (\emph{i.e.} cells for which cellular division is balanced by apoptosis) which is invaded by proliferating cancer cells. We group healthy cells into a population labelled by the index $H$, while cancer cells are grouped into a population labelled by the index $C$. We focus on a biological scenario whereby cells undergo pressure-driven movement (\emph{i.e.} they tend to move toward regions where they feel less compressed)~\cite{ambrosi2002closure}.

We represent each cell as an agent that occupies a position on a lattice and model cell movement and proliferation according to a set of simple rules that result in a discrete-time branching random walk. For ease of presentation, we let cells be arranged along the real line $\mathbb{R}$. 

We discretise the time variable $t \in \mathbb{R}_{\geq 0}$ and the space variable $x \in \mathbb{R}$ as $t_{k} = k \tau$ with $k\in\mathbb{N}_{0}$ and $x_{i} = i  \chi$ with $i \in \mathbb{Z}$, respectively, where $0<\tau, \chi \ll 1$. We let the dependent variables $N^{k}_{Hi}\in\mathbb{N}_0$ and $N^{k}_{Ci}\in\mathbb{N}_0$ model, respectively, the number of healthy cells and cancer cells on the lattice site $i$ at the time-step $k$, and we define the density of cells in population $h \in \{H,C\}$ and the total cell density, respectively, as
\begin{equation}
\label{e:n}
\rho_h(t_k,x_{i})= \rho^k_{hi} := N^{k}_{hi} \,  \chi^{-1} \; \text{ and } \; \rho(t_k,x_{i})= \rho^k_{i} := \rho^k_{Hi} + \rho^k_{Ci}.
\end{equation}
Moreover, for each lattice site $i$ and time-step $k$ we assume the cell pressure $p(t_k,x_{i})= p^k_{i}$ to be given by a barotropic relation $p^k_{i} \equiv \Pi(\rho^k_{i})$, where $\Pi$ is a function of the total cell density that satisfies the following conditions
\beq
\Pi(0) = 0,  \quad \frac{\rmd \Pi}{\rmd \rho}>0 \; \text{ for } \; \rho>0.
\label{as:p}
\eeq

The dynamics of healthy and cancer cells are governed by the following rules.

\paragraph{Mathematical modelling of cancer cell proliferation.} 
We allow every cancer cell to divide, die or remain quiescent with probabilities that depend on the cell pressure, and we assume that a dividing cell is replaced by two identical progeny  cells that are placed on the original lattice site of the parent cell. Focussing on the case of pressure-limited cell proliferation~\cite{basan2009homeostatic}, we let the net cell division rate $G$ be a function of the cell pressure $p$ that satisfies the following assumptions
\beq
\frac{\rmd G}{\rmd p} < 0, \quad G(P) =0 \quad \text{with} \quad P \in \mathbb{R}_{>0}.
\label{as:G}
\eeq
In~\eqref{as:G}, the parameter $P$ represents the pressure at which cell division is exactly balanced by cell death. At the $k^{th}$ time-step, a focal cancer cell on the lattice site $i$ can divide with probability 
\begin{equation}
\label{e:div}
\tau  \, G(p^{k}_{i})_+ \quad \mbox{where} \quad G(p^{k}_{i})_+ = \max\left(0,G(p^{k}_{i})\right)
\end{equation}
or die with probability 
\begin{equation}
\label{e:apo}
\tau  \, G(p^{k}_{i})_-  \quad \mbox{where} \quad G(p^{k}_{i})_- = - \min\left(0,G(p^{k}_{i})\right)
\end{equation}
or remain quiescent with probability 
\begin{equation}
\label{e:qui}
1 - \left(\tau  \, G(p^{k}_{i})_+ + \tau \, G(p^{k}_{i})_-\right) = 1 - \tau |G(p^{k}_{i})|.
\end{equation}
We assume the time-step $\tau$ to be sufficiently small so that the quantities~\eqref{e:div}-\eqref{e:qui} are all between 0 and 1. Under assumptions~\eqref{as:G}, the probabilities defined via \eqref{e:div}-\eqref{e:qui} are such that if $p^{k}_{i} > P$ then every cell on the $i^{th}$ lattice site can only die or remain quiescent at the $k^{th}$ time-step. Therefore, we have that
\begin{equation}
\label{eq:UBdisc} 
p^k_{i} \leq \overline{p} \; \mbox{ for all } \; (k,i) \in \mathbb{N}_0 \times \mathbb{Z}, \; \text{ with } \; \overline{p} = \max \left(\max_{i \in \mathbb{Z}} p^0_i, P \right).
\end{equation}

\paragraph{Mathematical modelling of cell movement.} We model pressure-driven cell movement by letting healthy and cancer cells move down pressure gradients according to a biased random walk whereby the movement probabilities depend on the difference between the pressure at the site occupied by a cell and the pressure at the neighbouring sites. In particular, for a focal cell of type $h \in \{H,C\}$ on the lattice site $i$ at the time-step $k$, we define the probability of moving to the lattice site $i-1$ (\emph{i.e.} the probability of moving left) as
\begin{equation}
\label{e:left}
J^{{\rm L}}_h(p^{k}_{i}-p^{k}_{i-1}) =  \nu_h \frac{(p^{k}_{i}-p^{k}_{i-1})_{+}}{2 \, \overline{p}}
\end{equation}
where $(p^{k}_{i}-p^{k}_{i-1})_{+} = \max\left(0,p^{k}_{i}-p^{k}_{i-1}\right)$, the probability of moving to the lattice site $i+1$ (\emph{i.e.} the probability of moving right) as
\begin{equation}
\label{e:right}
J^{{\rm R}}_h(p^{k}_{i}-p^{k}_{i+1}) = \nu_h \frac{(p^{k}_{i}-p^{k}_{i+1})_{+}}{2 \, \overline{p}} 
\end{equation}
where $(p^{k}_{i}-p^{k}_{i+1})_{+} = \max\left(0,p^{k}_{i}-p^{k}_{i+1}\right)$, and the probability of remaining stationary on the lattice site $i$ as 
\begin{equation}
\label{e:stay}
1 - J^{{\rm L}}_h(p^{k}_{i}-p^{k}_{i-1}) - J^{{\rm R}}_h(p^{k}_{i}-p^{k}_{i+1}).
\end{equation}
Here, the coefficient $0 < \nu_h \leq 1$ is directly proportional to the mobility of cells in population $h$ and the parameter $\overline{p}$ is defined via~\eqref{eq:UBdisc}. The a priori estimate~\eqref{eq:UBdisc} ensures that the quantities~\eqref{e:left}-\eqref{e:stay} are all between 0 and 1. 

\subsection{Corresponding continuum model and properties of its solutions}
\label{contmod2}
Using the formal method presented in~\cite{chaplain2019bridging} and letting $\tau \to 0$ and $\chi \to 0$ in such a way that
\beq
\label{asymat}
\frac{\nu_h \chi^{2}}{2 \, \overline{p} \, \tau} \rightarrow \mu_h \in \mathbb{R}_{>0} \quad \text{for} \quad h \in \{H,C\},
\eeq
where the parameter $\mu_h$ is the mobility of cells in population $h$, it is possible to show that the deterministic continuum counterpart of the stochastic individual-based model presented in Section~\ref{sec:discreteTumInv} is given by the following system of conservation equations for the density of healthy cells $\rho_H(t,x) \geq 0$ and the density of cancer cells $\rho_C(t,x) \geq 0$:
\beq
\label{eq:pde3}
\left\{
\begin{array}{ll}
\partial_t \rho_H - \mu_H \, \partial_x \left(\rho_H \, \partial_x p \right) = 0, 
\\\\
\partial_t \rho_C - \mu_C \, \partial_x \left(\rho_C \, \partial_x p \right) = G(p) \, \rho_C,
\end{array}
\right.
\quad 
(t,x) \in (0, \infty) \times \mathbb{R}.
\eeq
In the system of nonlinear PDEs~\eqref{eq:pde3}, the cell pressure $p(t,x)$ is defined as a function of the total cell density $\rho(t,x) := \rho_H(t,x) + \rho_C(t,x)$ via a barotropic relation $\Pi(\rho)$ and the function $G(p)$ satisfies assumptions~\eqref{as:G}.

Considering the following simplified barotropic relation proposed by Perthame {\it et al.}~\cite{perthame2014hele}
\beq
\label{def:p}
\Pi(\rho) := K_{\gamma} \, \rho^{\gamma} \quad \text{with} \quad \gamma > 1 \quad \text{and} \quad K_{\gamma}>0,
\eeq
where $\gamma$ provides a measure of the stiffness of the barotropic relation and $K_{\gamma}$ is a scale factor, Lorenzi \emph{et al.}~\cite{lorenzi2017interfaces} proved the existence of travelling-wave solutions to~\eqref{eq:pde3} of the form
$$
\rho_H(t,x) = \rho_H(z), \quad \rho_C(t,x) = \rho_C(z), \quad z = x - \sigma \, t, \quad \sigma > 0
$$
that satisfy the conditions
\beq
\label{ass:TW2a}
\rho_C(z)\left\{
\begin{array}{ll}
> 0, \quad \text{for } z < 0,
\\\\
= 0 , \quad \text{for } z \geq 0,
\end{array}
\right.
\qquad
\rho_H(z)\left\{
\begin{array}{ll}
= 0 , \quad \text{for } z < 0,
\\\\
> 0, \quad \text{for } z \in [0, \ell),
\\\\
= 0 , \quad \text{for } z \geq \ell,
\end{array}
\right.
\eeq
for some $\ell > 0$, along with the asymptotic condition 
\beq
\label{ass:TW2b}
\rho_C(z) \; \xrightarrow[z  \rightarrow -\infty]{} \Pi^{-1}(P).
\eeq
Conditions~\eqref{ass:TW2a} correspond to a scenario in which healthy and cancer cells do not mix and are separated by a sharp interface in $z=0$. The travelling-wave solutions $\rho_C(z)$ and $\rho_H(z)$ are non-negative and non-increasing, and the corresponding cell pressure $p(z)$ is continuous, non-increasing and has a kink at $z=0$ (\emph{i.e.} at the interface between the two cell populations) with
\begin{equation}
\label{eq:ppjump}
\sgn\big(p'(0^+) - p'(0^-)\big) = \sgn (\mu_H-\mu_C).
\end{equation}
Finally, it was numerically shown that such travelling-wave solutions can become unstable in the case where $\mu_C > \mu_H$, which leads to the occurrence of spatial mixing between cancer and healthy cells. 

\subsection{Quantitative comparison between discrete and continuum models}
\label{secnum2n}
In this section, we compare the outcomes of numerical simulations of the stochastic individual-based model presented in Section~\ref{sec:discreteTumInv} with numerical solutions of the deterministic continuum model given by~\eqref{eq:pde3}, and we briefly discuss their biological implications. The results obtained indicate excellent agreement between the simulation results for the individual-based model, the numerical solutions of the continuum model and the travelling-wave results summarised in Section~\ref{contmod2}. 

\subsubsection{Numerical methods and set-up of numerical simulations}
To construct numerical solutions of~\eqref{eq:pde3}, we use a uniform discretisation consisting of 1001 points on the interval $[0,100]$ as the computational domain of the independent variable $x$. We employ a finite volume method based on a time-splitting between the conservative and non-conservative parts to solve numerically~\eqref{eq:pde3} subject to no-flux boundary conditions and to the following initial conditions
$$
\rho^0_{C}(x)  := A_{C} \exp{\left(-b_{C} \, x^{2}\right)} 
$$
and
$$
\rho^0_{H}(x):=\left\{
\begin{array}{ll}
0 , \quad \text{for } x \in [0,13],
\\\\
A_{H} \exp{\left(-b_{H} (x-14)^{2}\right)}, \quad \text{for } x \in (13,29),
\\\\
0 , \quad \text{for } x \in [29,100],
\end{array}
\right.
$$
where
$$
A_{C}=1.25\times10^{4}, \quad  b_{C}=6 \times 10^{-2}, \quad A_{H}=2.5\times10^{4} \quad \mbox{and} \quad b_{H}=6\times 10^{-3}.
$$
For the conservative parts, transport terms are approximated through an upwind scheme whereby the cell edge states are calculated by means of a high-order extrapolation procedure~\cite{leveque2002finite}, while the forward Euler method is used to approximate the non-conservative parts. We use the following definition of the net cell division rate $G$
$$
G(p) := \frac{1}{2 \, \pi} \, \arctan(\theta \, (P-p)), \quad \theta = 4 \times 10^{-5}, \quad P=5 \times 10^9,
$$
and define the cell pressure via the barotropic relation~\eqref{def:p} with 
$$
K_{\gamma} := \dfrac{\gamma+1}{\gamma} \quad \text{and} \quad \gamma:=2.
$$

An analogous set-up is used to carry out computational simulations of the discrete model with $\tau=10^{-3}$ and $\chi=0.1$. Given the values of $\nu_H$ and $\nu_C$ (see captions of Figure~\ref{fig3} and~\ref{fig4} for parameter values), the values of $\mu_H$ and $\mu_C$ are defined in such a way that condition~\eqref{asymat} is met. At each time-step, we follow the procedures summarised hereafter to simulate cancer cell proliferation and movement of healthy and cancer cells. Numerical simulations are performed in {\sc Matlab} and all random numbers mentioned below are real numbers drawn from the standard uniform distribution on the interval $(0,1)$ using the built-in function {\sc rand}. 
\\\\
{\bf Numerical simulation of cancer cell proliferation} The number of cells and the cell pressure are computed, and the probabilities of division, death and quiescence of the cancer cells are evaluated via~\eqref{e:div}-\eqref{e:qui} for every lattice site. For each cell, a random number is generated and the cells' fate is determined by comparing this number with the probabilities of division, death and quiescence at the cell lattice site.
\\\\
{\bf Numerical simulation of cell movement} The cell pressure is computed and used to evaluate the probabilities of moving left, moving right and remaining stationary via~\eqref{e:left}-\eqref{e:stay} for every lattice site. For each cell, a random number is generated and it is determined whether or not the cell will move by comparing this number with the value of the sum of the probability of moving left and the probability of moving right. If a cell moves, a new random number is generated and the cell move either onto the lattice site to the left or onto the lattice site to the right of its current lattice site based on a comparison between the random number and the values of the probability of moving left and the probability of moving right. No-flux boundary conditions are implemented by letting the attempted movement of a cell be aborted if it requires moving out of the computational domain.

\subsubsection{Main results}
\begin{figure}[h!]
\centering
\includegraphics[width=0.85\textwidth]{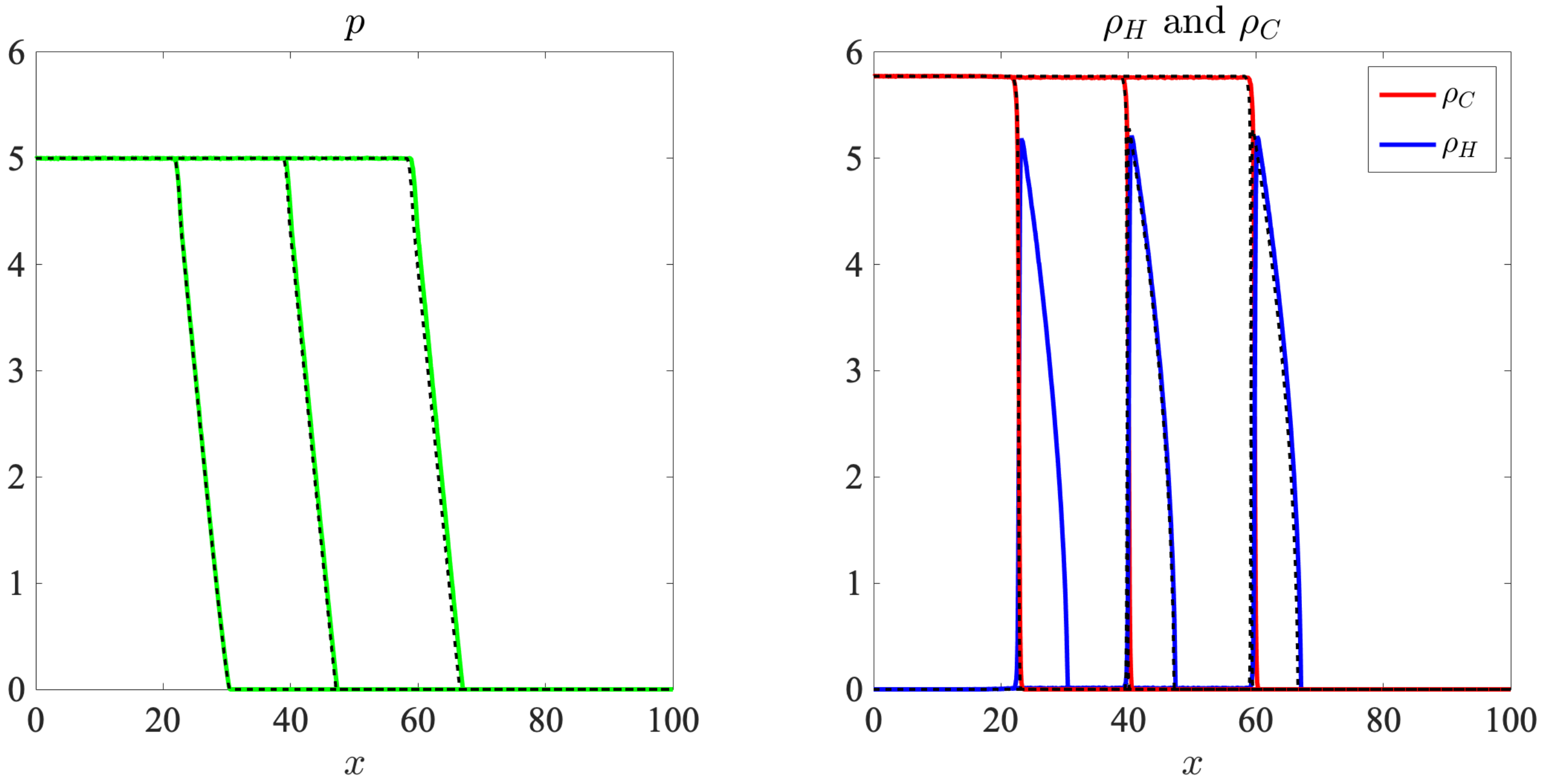}
\caption{{\bf Dynamics of the pressure and the cell densities for $\nu_H \geq \nu_C$.} Comparison between the results of numerical simulations of the individual-based model (solid lines) and the numerical solutions of~\eqref{eq:pde3} (dashed lines), for $\nu_H \geq \nu_C$ (\emph{i.e.} $\mu_H \geq \mu_C$). The left panel displays the cell pressure at three successive time instants -- \emph{i.e.} $t=5 \times 10^4$ (left curve), $t=12 \times 10^4$ (central curve) and $t=20 \times 10^4$ (right curve) -- while the right panel displays the corresponding densities of cancer cells (red lines) and healthy cells (blue lines). Values of the cell pressure are in units of $10^9$, while values of the cell densities are in units of $10^4$. Simulations were carried out with $\nu_H=0.2$ and $\nu_C=0.1$, and defining the values of the parameters $\mu_H$ and $\mu_C$ in such a way that condition~\eqref{asymat} was met.}
\label{fig3}
\end{figure}

\begin{figure}[h!]
\centering
\includegraphics[width=0.85\textwidth]{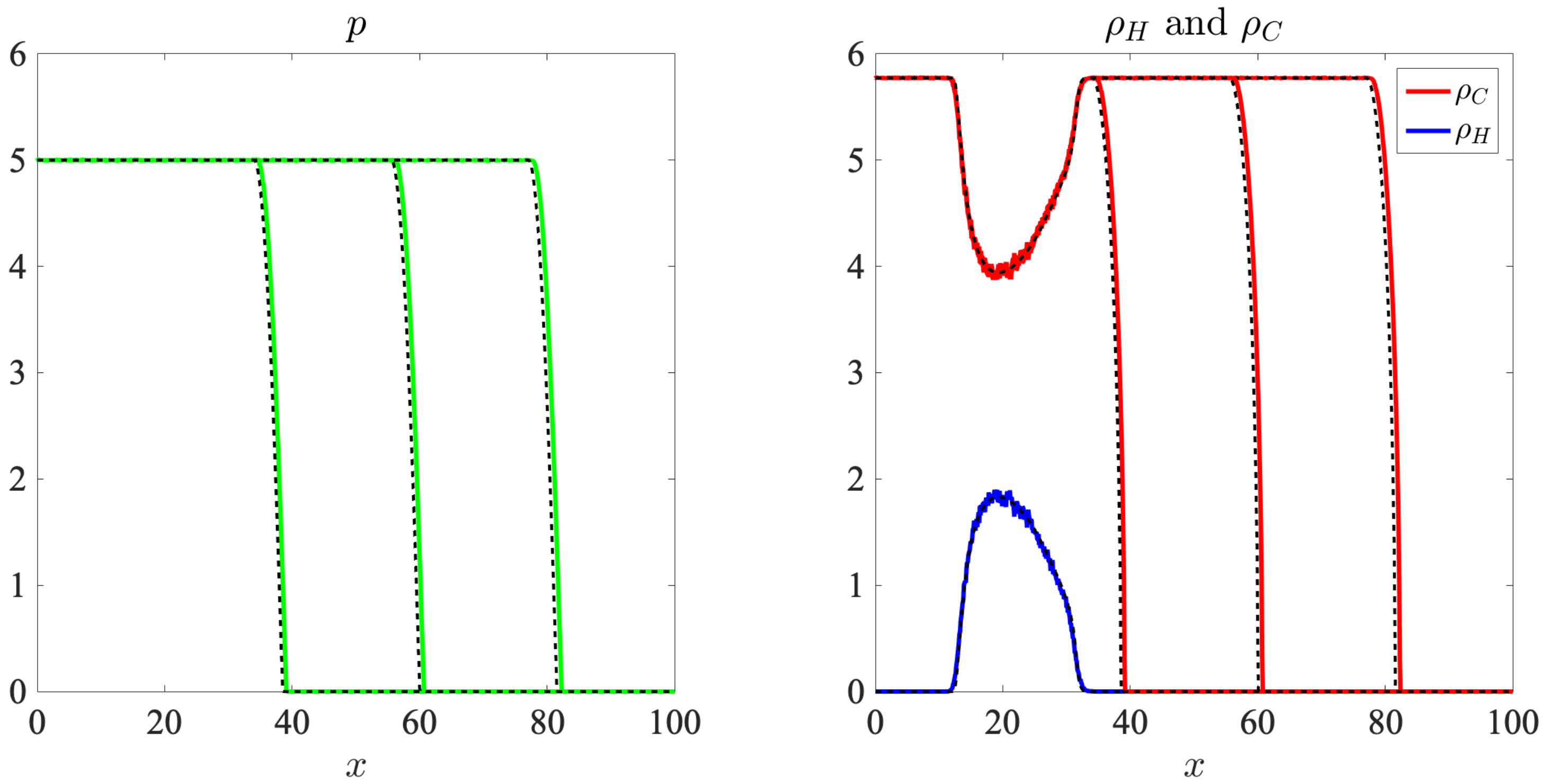}
\caption{{\bf Dynamics of the pressure and the cell densities for $\nu_C > \nu_H$.} Comparison between the results of numerical simulations of the individual-based model (solid lines) and the numerical solutions of~\eqref{eq:pde3} (dashed lines), for $\nu_C > \nu_H$ (\emph{i.e.} $\mu_C > \mu_H$). The left panel displays the cell pressure at three successive time instants  -- \emph{i.e.} $t=5 \times 10^4$ (left curve), $t=7.5 \times 10^4$ (central curve) and $t=10 \times 10^4$ (right curve) -- while the right panel displays the corresponding densities of cancer cells (red lines) and healthy cells (blue lines). Values of the cell pressure are in units of $10^9$, while values of the cell densities are in units of $10^4$. Simulations were carried out with $\nu_H=0.1$ and $\nu_C=0.2$, and defining the values of the parameters $\mu_H$ and $\mu_C$ in such a way that condition~\eqref{asymat} was met.}
\label{fig4}
\end{figure}

The plots in Figures~\ref{fig3} and~\ref{fig4} show sample dynamics of the cell pressure (left panel, solid line) and the cell densities (right panel, solid lines) computed from numerical simulations of the individual-based model and corresponding numerical solutions of~\eqref{eq:pde3} (dashed lines). 

In agreement with the travelling-wave results summarised in Section~\ref{contmod2}, the numerical results presented in Figure~\ref{fig3} show that if $\nu_H \geq \nu_C$ (\emph{i.e.} $\mu_H \geq \mu_C$) spatial segregation occurs and the two cell populations remain separated by a sharp interface. Healthy cells stay ahead of cancer cells and cell densities are non-increasing. The pressure itself is continuous across the interface between the two cell populations, whereas its first derivative jumps from a smaller negative value to a larger negative value -- \emph{i.e.} the sign of the jump coincides with $\sgn(\mu_H - \mu_C)$ (see the jump condition~\eqref{eq:ppjump}). 

\newpage
On the other hand, the numerical results displayed in Figure~\ref{fig4} indicate that when $\nu_C > \nu_H$ (\emph{i.e.} $\mu_C > \mu_H$) spatial mixing between cancer cells and healthy cells occurs. This is consistent with the fact that the travelling-wave solutions discussed in Section~\ref{contmod2} are expected to be unstable in the case where $\mu_C > \mu_H$. 
\\\\
{\bf Biological implications in brief} These results communicate the biological notion that larger values of the mobility of cancer cells facilitate the formation of infiltrating, malignant patterns of invasion, whereas non-infiltrating patterns of invasion are to be expected in the scenarios where the mobility of healthy cells is higher than that of cancer cells.

\section{Conclusions and possible developments of the models}
\label{Section4}
We presented a stochastic individual-based model for the evolutionary dynamics of cancer cells under cytotoxic therapy and a stochastic individual-based model for the mechanical interaction between healthy and cancer cells during tumour growth. We showed that the continuum counterpart of the former model is given by a non-local PDE for the cell population density, while the continuum counterpart of the latter model comprises a system of coupled nonlinear PDEs for the densities of cancer and healthy cells. We discussed some key properties of the solutions of the PDEs and compared the results of numerical simulations of the individual-based models with numerical solutions of the corresponding PDEs. We found excellent quantitative agreement between the numerical simulation results of discrete and continuum models and briefly discussed their biological implications. We conclude with an outlook on possible developments of these models.
\\\\
\noindent{\bf Possible developments of the model for the evolutionary dynamics of cancer cells under cytotoxic therapy} We let the the concentration of cytotoxic agent remain constant in time. However, the stochastic individual-based model presented here, as well as the corresponding deterministic continuum model, could be easily adapted to the case where the dose of cytotoxic agent varies over time. In this regard, it would be interesting to investigate whether the delivery schedules for the cytotoxic agent obtained through numerical optimal control of the non-local PDE~\cite{almeida2018evolution,olivier2017combination} would remain optimal also for the individual-based model. Another track to follow might be to investigate the effect of stress-induced phenotypic variations triggered by the selective pressure that cytotoxic agents exert on cancer cells~\cite{chisholm2015emergence}. An additional development of the model presented here would be to include a spatial structure, for instance by embedding the cancer cells in the geometry of a solid tumour, and to take explicitly into account the effect of spatial interactions between cancer cells, therapeutic agents and other abiotic factors, such as oxygen and glucose~\cite{lorenzi2018role,lorz2015modeling,villa2019}. In this case, the resulting individual-based model would be integrated with a system of PDEs modelling the dynamics of the abiotic factors, thus leading to a hybrid model~\cite{anderson1998continuous,bouchnita2017hybrid,bouchnita2016bone,burgess2017examining,burgess2016dynamical,eymard2016mathematical,franssen2018mathematical,hamis2018does,kurbatova2011hybrid,schofield2002mathematical,schofield2005dynamic}. 
\\\\
{\bf Possible developments of the model for the mechanical interaction between healthy and cancer cells during tumour growth} We focussed on the case of one spatial dimension and considered barotropic relations that satisfy~\eqref{as:p}. However, the stochastic individual-based model presented here, as well as the corresponding deterministic continuum model, could be adapted to higher spatial dimensions and more realistic barotropic relations. In this regard, it would be interesting to use such a stochastic individual-based model to further investigate the formation of finger-like patterns observed for the system of equations~\eqref{eq:pde3} posed on a two dimensional spatial domain~\cite{lorenzi2017interfaces}. These spatial patterns resemble infiltrating patterns of cancer-cell invasion commonly observed in breast tumours~\cite{wang2012adipose}. An additional development would be to compare the results of numerical simulations of this model with those obtained from equivalent discrete models defined on irregular lattices, as well as to investigate how the modelling approach considered here could be related to off-lattice models of growing cell populations~\cite{drasdo2005coarse,van2015simulating}. 

\bibliographystyle{siam}
\bibliography{biblio}

\end{document}